\documentclass[prl,aps,twocolumn,superscriptaddress,showpacs,floatfix]{revtex4}
\usepackage{amsmath}
\usepackage{amsbsy}
\usepackage{amssymb}
\usepackage[dvips]{graphicx}
\usepackage{epsfig}
\usepackage{color}
\usepackage[normalem]{ulem}
\usepackage{subfigure}

\usepackage{psfrag}

\def\bse{\begin{subequations}}
\def\ese{\end{subequations}}

\begin{document}

\title{Non equilibrium phase transition with gravitational-like
  interaction in a cloud of cold atoms}%

\author{J. Barr\'e}
\affiliation{Laboratoire J.-A. Dieudonn\'e, Universit\'e de Nice Sophia-Antipolis, CNRS, 06109 Nice, France.}
\author{B. Marcos}
\affiliation{Laboratoire J.-A. Dieudonn\'e, Universit\'e de Nice Sophia-Antipolis, CNRS, 06109 Nice, France.}
\affiliation{Instituto de F\'{\i}sica, Universidade Federal do Rio Grande do
 Sul, Caixa Postal 15051, CEP 91501-970, Porto Alegre, RS, Brazil.}
\author {D. Wilkowski}
\affiliation{Institut Non Lin\'eaire de Nice, Universit\'e de Nice Sophia-Antipolis, CNRS, 06560 Valbonne, France.}
\affiliation{Centre for Quantum Technologies, National University of Singapore , 117543 Singapore, Singapore.}
\affiliation{PAP, School of Physical and Mathematical Sciences, Nanyang Technological University, 637371 Singapore, Singapore.}

\date{\today{}}
\begin{abstract}

  We propose to use a cloud of laser cooled atoms in a quasi two
  dimensional trap to investigate a non equilibrium collapse phase
  transition in presence of gravitational-like interaction. Using
  theoretical arguments and numerical simulations, we show that, like
  in two dimensional gravity, a transition to a collapsed state occurs
  below a critical temperature. In addition and as a signature of the
  non equilibrium nature of the system, persistent particles currents,
  dramatically increasing close to the phase transition, are observed.

\end{abstract}
%

\maketitle

Non equilibrium phase transitions have been extensively studied over
the years both for basic understanding and potential applications
\cite{RevModPhys.76.663}. Among the numerous examples of non
equilibrium phase transitions, one can quote: direct percolation
\cite{coniglio1981thermal}, infection spreading \cite{Hinrichsen},
geophysical flows \cite{Berhanu}, complex plasmas~\cite{Sutterlin},
surfaces \cite{Barabasi} and nanowire growing \cite{krogstrup} and
traffic jams \cite{Wolf}.

For equilibrium phenomena, a systematic approach exists, and powerful
tools such as the renormalization group have been developed. In
contrast, and despite important progresses in some cases (see
\cite{Henkel} for a textbook account) there is no such general
framework so far for non equilibrium phase transitions
\cite{RevModPhys.76.663}. This is an outstanding open problem of
statistical physics, since most biological, chemical and physical
systems encountered in nature as well as social phenomena are in non
equilibrium states.

In this letter, we study a non equilibrium phase transition driven by
an effective gravitational-like interaction, which does not derive
from a potential, in a quasi two dimensional (2D) cloud of laser
cooled atoms. At equilibrium, inter-particle long-range
interactions are at the origin of dramatic collective effects, such as
gravothermal catastrophe or isothermal collapse in self-gravitating
systems ~\cite{Antonov1961,LyndenBell68,Kiessling89}, and
negative specific
heat~\cite{LyndenBell68,Thirring70}. Systems of Brownian self-gravitating particles in 2D undergo a collapse phase
transition, in the sense that the density
diverges in finite time below a critical temperature~\cite{Chavanis2002a,Chavanis2002}. For our non equilibrium system, we find a
similar behavior. In addition, we
observe, as a direct signature of the presence of non equilibrium
state, persistent currents which are rapidly growing close to the
transition. Those particularities are explained throughout the letter.

\begin{figure}
\includegraphics[width=0.3\textwidth]{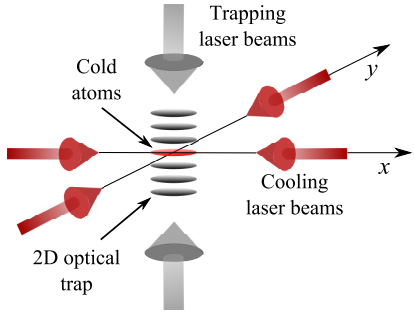}
\caption{Schematic view of the experimental setup. Atoms are confined
  in strongly anisotropic, quasi two dimensional, optical dipole
  traps: they form one or several pancake shaped clouds. Two pairs of
  orthogonal contra-propagating lasers beams are in quasi-resonance
  with an atomic transition providing cooling and shadow forces in the
  x-y plane.}
\label{fig1}
\end{figure}

\emph{System-} The starting point of our studies is a simple
experimental setup where a cooled atomic gas is located in the x-y
plane into one or few two dimensional traps made of a far off-detuned
stationary laser beam (see Fig. \ref{fig1}). The dynamic along the
perpendicular axis of the traps is frozen due to a strong
confinement. In the x-y plane, the gas is in interaction with two
orthogonal contra-propagating pairs of red detuned laser beams
providing laser cooling. Hence, the laser beams can be seen as a
thermal bath at temperature $T$. In addition, these
  cooling beams are absorbed by the gas (shadow effect), leading, in
the weak absorption limit, to a gravitational-like interaction
in the xy plane, along the axis of the beams
\cite{Dalibard1988}. This interaction was experimentally demonstrated
in a one dimensional system \cite{Chalony2013}. The strength of the
interaction can be tuned changing the intensity or/and the frequency
detuning of the laser beams (see Eq. \eqref{force-constant}).
Since the interaction force satisfies the Poisson
  equation, this system might at first sight be viewed as a tabletop
realization of a 2D gravitational-like system in canonical
equilibrium. However, the spatial configuration of the laser beams
does not preserve the rotational symmetry around the z axis.  Hence, the laser-induced long range force does not derive
anymore from a potential, and the system is fundamentally in a non
equilibrium state.

\emph{Modeling}- We describe one cloud of cold atoms by a two
dimensional phase space density $f(\vec{r},\vec{v},t)$, using the
shorthands $\vec{r}=(x,y)$, $\vec{v}=(v_x,v_y)$.  We make the
reasonable assumption that cooling can be modeled by a linear friction
$-\gamma \vec{v}$, and use a constant velocity diffusion coefficient
$D$ to take into account the velocity recoils due to random absorption
and fluorescence emission of photons by the atoms.  We use the
following standard expression for the force $\vec{F}=(F_x,F_y)$ for
the shadow effect, which relies on a weak absorption limit
\cite{Dalibard1988}: \bse
\label{def-force}
\begin{align}
F_x[\rho](x,y) &= -C \int \mbox{sgn}(x-x') \rho(x',y)~dx'  \label{eq:Fx}\\
F_y[\rho](x,y) &= -C \int \mbox{sgn}(y-y') \rho(x,y')~dy' \label{eq:Fy}
\end{align}
\ese where $\rho(x,y)$ is the surface density of atoms, normalized to
$1$; $C$ is a constant characterizing the intensity of the force,
which can be computed after integration over the transverse
direction~\cite{Chalony2013}:
\begin{equation}
\label{force-constant}
  C=  \frac{\hbar k \Gamma}{2}\frac{I_0}{I_s} \frac{N}{(1+4\bar{\delta}^2)^2}\frac{\sigma_0}{2\sqrt{\pi}L_{\perp}}.
\end{equation}
In this expression, $k$ is the wave number, $\Gamma$ the width of the
atomic transition, $\bar{\delta}$ the normalized frequency detuning,
$N$ the number of trapped atoms, $\sigma_0=6\pi/k^2$ the resonant
photon absorption cross-section, $L_{\perp}$ the transverse size of
the cloud, $I_0$ the incident laser intensity and $I_s$ the saturation
intensity.  Note that the shadow force \eqref{def-force} verifies the
same Poisson equation as gravity $\vec{\nabla}\cdot \vec{F} \propto
-\rho$, but, contrary to gravity, does not derive from a potential,
i.e., $\vec{\nabla}\times \vec{F}\neq \vec 0$.

The optical dipole traps in the x-y plane are well approximated by
harmonic traps at frequency $\omega_0$ and will be modeled
accordingly. Moreover, usual experimental configurations correspond to
the overdamped regime, i.e. $\omega_0\ll\gamma$. So the velocity
distribution quickly relaxes to a Gaussian, and the surface density
$\rho(x,y,t)$ evolves according to a non-linear Smoluchowski equation
\begin{equation}
  \frac{\partial \rho}{\partial t} = \vec{\nabla}\cdot \left[ \frac{\omega_0^2}{\gamma} \vec{r} \rho
    + \frac{1}{m\gamma}\vec{F}[\rho] \rho +\frac{k_BT}{m\gamma} \vec{\nabla}\rho \right]
\label{eq:smol}
\end{equation}
where $m$ is the atomic mass, and the temperature is determined by the relation
  $k_BT/m=D/\gamma$.

  Rescaling time and space as $\tilde{t}=(\omega_0^2/\gamma)t$,
  $(x,y)=(L\tilde{x},L\tilde{y})$ with $L=\sqrt{C/(m\omega_0^2)}$,
  Eq.~\eqref{eq:smol} becomes (dropping the \,$\tilde{}$\, for
  convenience):
\begin{equation}
\frac{\partial \rho}{\partial t} = \vec{\nabla}\cdot\left[ \vec{r} \rho
+ \vec{F}[\rho] \rho +\Theta \vec{\nabla}\rho \right]
\label{eq:smol2}
\end{equation}
This equation is the starting point of our
theoretical analysis.

{\it Model analysis-} The physics is governed by a single
dimensionless parameter
\[
\Theta = \frac{k_BT}{C}.
\]
The above Eq.~(\ref{eq:smol2}) is similar to the Smoluchowski-Poisson
system describing self-gravitating Brownian particles, or the
parabolic-elliptic Keller-Segel model used in chemotaxy
theory\cite{KellerSegel,Chavanis1}. However, the force does not derive
from a potential.  It is well-known that if the temperature is small
enough, a solution to the Smoluchowski-Poisson equation blows up in
finite time and forms a Dirac peak. This behavior can also be seen as
a phase transition in the canonical ensemble. It is natural to ask if
the same phenomenology holds for Eq.~(\ref{eq:smol2}), the non
potential generalization of the Smoluchowski-Poisson equation.

To get insight in the behavior of Eq.~(\ref{eq:smol2}), we compute the
time evolution of $S(t)= -\int \rho \ln \rho$. Note that the
  formation of a Dirac peak corresponds to $S(t)\to -\infty$. Using
Eq.~\eqref{eq:smol2}, $\vec{\nabla}\cdot \vec{F} = -4\rho$ and after
integrating by parts, we get
\begin{equation}
\dot{S}(t) = -2 -4\int \rho^2d^2\vec{r} +\Theta \int \frac{|\nabla \rho|^2}{\rho}d^2\vec{r}~.
\end{equation}
Writing now $\rho=\sqrt{u}$ and using the functional inequality, valid
in 2D \cite{Weinstein82}: $\int \! u^4\leq a \int\! |\nabla u|^2
\times \int\! u^2$, we have
\begin{equation}
\dot{S}(t) \geq -2 + (\Theta -a)  \int \!\frac{|\nabla \rho|^2}{\rho}d^2\vec{r}
\end{equation}
where $a$ is a constant, known numerically: $a\simeq 0.171$. If
$\Theta>a$, the second term in the right hand side of the inequality
is positive and dominates over the first term above a certain spatial
density of the cloud. It ensures that $S(t)$ cannot decrease without
bound: collapse is excluded. On the other hand, for $\Theta<a$,
collapse becomes possible, even though of course this argument cannot
prove that it happens. If it happens, we should not expect either
$0.171$ to be an accurate estimate of the critical parameter
$\Theta_c$, but an indication for the behavior of
Eq.~\eqref{eq:smol2}. Indeed, as we describe below, we numerically
find indications of a collapse transition at a lower $\Theta$, namely
$\Theta\simeq 0.12-0.15$.

{\it Numerical simulations-} To simulate Eq.\eqref{eq:smol2},
we introduce the following stochastic particles approximation:
The position of particle $i$ is denoted by $\vec{r}_i=(x_i,y_i)$, and the
dynamical equations are:
\bse
\label{eq-motion-sim}
\begin{align}
\dot{x}_i &=  -x_i+F_{i,x}+\sqrt{2\Theta}\eta_{i,x}(t) \\
\dot{y}_i &=  -y_i+F_{i,y}+\sqrt{2\Theta}\eta_{i,y}(t),
\end{align}
\ese
where the $\eta_{i,.}$ are independent gaussian white noises.
To define $F_{i,x}$ and $F_{i,y}$ numerically we introduce the spatial
scale $\sigma$. The force is then written as
\bse
\label{def-force-simu}
\begin{align}
F_{i,x} &= -C \sum_{j\ne i} \mbox{sgn}(x_i-x_j) \delta_\sigma(y_i-y_j) \\
F_{i,y} &= -C \sum_{j\ne i} \mbox{sgn}(y_i-y_j) \delta_\sigma(x_i-x_j),
\end{align}
\ese where $\delta_\sigma(z)=1$ if $|z|<\sigma$ and zero otherwise. In
the limit $\sigma\to0$, Eqs.\eqref{def-force-simu} reduce to the
original definition of the force \eqref{def-force}. We expect to
correctly approximate Eq.\eqref{eq:smol2} when $n\sigma\gg 1$, where
$n$ is the number of particles. We integrate the equation of motions
using an Euler scheme. The force calculation is sped up using the
following procedure: space is discretized with cells of size $\sigma$
and particles are assigned to cells using the linked-list technique
(see e.g. \cite{knuth_68}); this is a $\mathcal O(n)$ operation which
does not involve any approximation.  Note that the numerical particles
should not be seen as direct representations of the atoms in the trap;
however, the spatial distribution of the numerical particles should
approximate the 2D spatial distribution of atoms described by
\eqref{eq:smol2}.

We performed a series of simulations varying initial conditions,
$\sigma$ and number of particles $n$, with $\Theta$ in the range
$[0.08,0.3]$. In order to keep the strength of the gravitation-like
interaction constant when changing the parameters, we keep the
quantity $Cn\sigma$ constant.  After a time $t\sim 1$, all the
simulations reach a stationary state, which we find to be essentially
independent of $\sigma$, $n$ (for sufficiently large $n$ and small
$\sigma$) and of the initial conditions.
\begin{figure}
\begin{center}
   \includegraphics[height=0.23\textwidth]{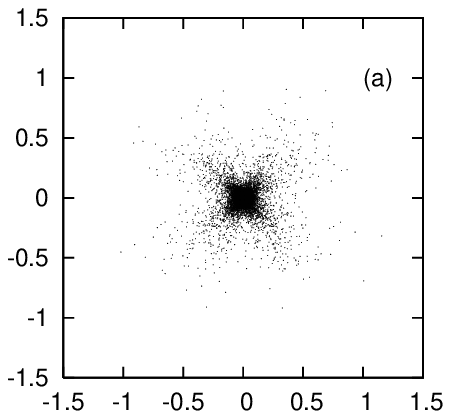}
 \includegraphics[height=0.23\textwidth]{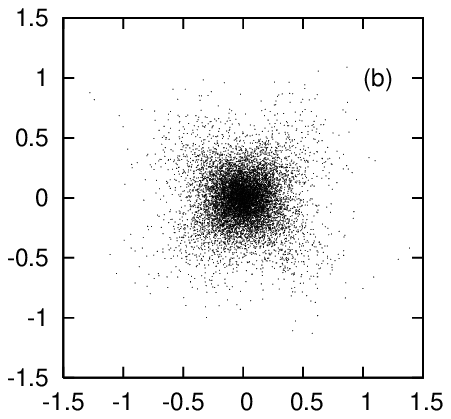}
 \caption{ Snapshot of the particles distribution in the stationary
   state of a system with $\sigma=10^{-2}$, $n=10^4$, a time step
   $\Delta t=10^{-5}$ and (a) $\Theta=0.14$ and (b) $\Theta=0.2$. The
   laser beams are along the axes of the figure.}\label{snap}
\end{center}
\end{figure}
In Fig.~\ref{snap} are shown snapshots of the particles distributions
in the stationary state at $\Theta=0.14$ and $\Theta=0.2$. They show a
cross-like structure along the diagonals, which is related to the
presence of currents as we will discuss latter on.  To look for a
phase transition toward a collapse phase, we plot the central spatial
density as a function of $\Theta$, as shown in
Fig.~\ref{fig_dens0}. We observe an abrupt increase in the density
when $\Theta$ is decreased, for $\Theta_c\approx 0.12-0.15$.
\begin{figure}
\begin{center}
  \psfrag{X}{$\Theta$}
  \psfrag{Y}{$\rho(r=0)$}
   \includegraphics[height=0.35\textwidth]{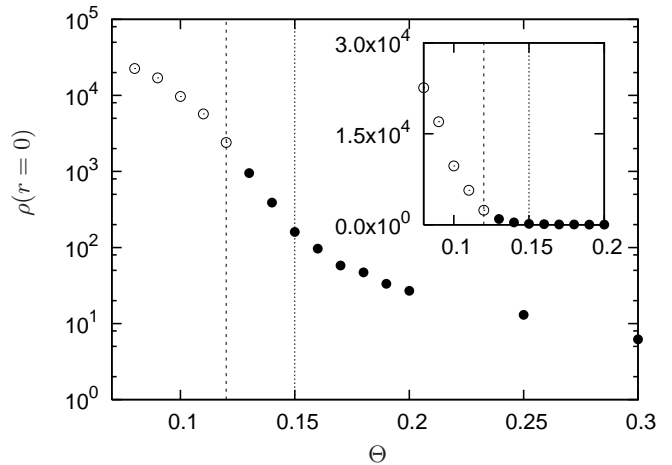}
   \caption{Density $\rho(r=0)$ in the center of the trap, as a
     function of the reduced temperature $\Theta$ for
     $\sigma=10^{-2}$, $n=10^4$ and a time step $\Delta
     t=10^{-5}$. Filled (empty) points correspond to simulations which
     (do not) have numerically converged with respect to the time step
     (see text). \emph{Inset}: Same quantity plotted in linear-linear
     scale. The two vertical dashed lines indicate the numerically
     estimated location of the transition
     $\Theta_c\approx0.12-0.15$. \label{fig_dens0} }
\end{center}
\end{figure}

Furthermore, we note that for all simulations with $\Theta >\Theta_c$
the asymptotic state is independent of the time step, suggesting a
convergence to a regular stationary state of \eqref{eq:smol2}. In
contrast, for $\Theta <\Theta_c$, we have been unable to reach a
stationary state independent of the time step. Hence the numerical
results for $\Theta < \Theta_c$ should be taken with caution, since no
convergence to a regular solution is achieved. Importantly, this lack
of convergence suggests as well that the limit system develops a
singularity, indicating the presence of the collapsed phase.

\begin{figure}
\begin{center}
  \psfrag{X}{$\Theta$}
  \psfrag{Y}{$\langle|\vec J|^2\rangle$}
   \includegraphics[height=0.35\textwidth]{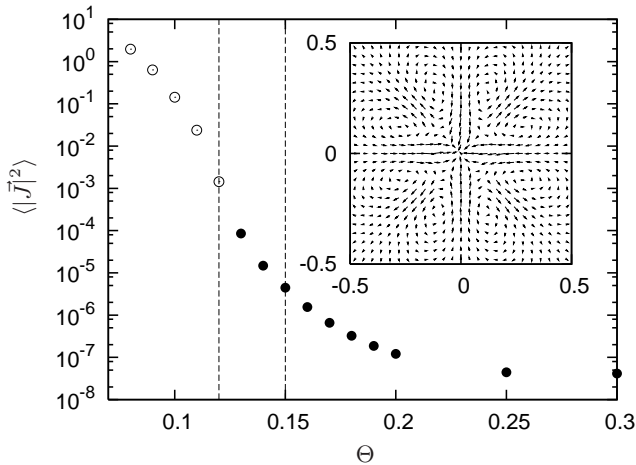}
   \caption{Spatially averaged square intensity of the currents as a
     function of $\Theta$ for $\sigma=10^{-2}$, $n=10^4$ and a time
     step $\Delta t=10^{-5}$. Filled (empty) points correspond to
     simulations which (do not) have numerically converged with
     respect to the time step (see text). The vertical dashed lines
     indicates the numerically estimated location of the transition
     region: $\Theta_c\approx0.12-0.15$. \emph{Inset}: Spatial
     distribution of current $\vec J$ in the stationary state; 
$\Theta=0.2$. The
     laser beams are along the axes of the figure.} \label{var_flux}
\end{center}
\end{figure}

A phase transition toward a collapsed phase at a finite temperature
parameter $\Theta$ makes our system similar to a 2D self-gravitating
gas of Brownian particles, for which the phase transition is predicted
at $\Theta_c =1/(2\pi)$ (see for example \cite{Chavanis2002}), i.e. at
a value slightly larger than the one numerically found here.  However,
in contrast with a 2D self-gravitating gas, the system is truly out of
equilibrium; this can be illustrated by computing the current $\vec J$
in the stationary state. Using Eq.~\eqref{eq:smol2}, we have $\vec J =
\vec{r} \rho + \vec{F}[\rho] \rho +\Theta \vec{\nabla}\rho$, with
$\vec{\nabla} \vec J =0$. For interactions deriving from a potential,
it is simple to show that, in thermal equilibrium, $\vec J = \vec
0$. In the present case $\vec J \ne \vec 0$, and the inset of
Fig.~\ref{var_flux} shows their spatial structure. As expected,
ingoing currents are along the laser beams where the long-range
interaction is maximal. Escaping channels from the trap center are
along the diagonals. This current structure explains the cross-like
shape found in the particle distribution (see Fig. \ref{snap}). In the
main part of the figure, we see that the current intensity strongly
increases as the transition region is approached from above. Like for
the central density discussed above, the computed current intensity in
the $\Theta<\Theta_c$ region should be taken with caution, since the
simulations results still depend on the time step.

{\it Possible experimental realization-} The experiment could be
performed following the scheme depicted in
ref. \cite{Chalony2013}. The starting point would be a Strontium88
gas, laser cooled in a magneto-optical trap operating on the narrow
$^1\!S_0\rightarrow\,^3\!P_1$ intercombination line at
$689\,\textrm{nm}$ \cite{chaneliere2008three}. The bare linewidth of
the transition is $\Gamma/2\pi = 7.5\,\textrm{kHz}$. Ultimately, the
gas is transferred into one or several 2D dipole traps made with a far
off-detuned high power standing optical wave located along the
vertical axis. We expect that the interactions between two parallel 2D
traps shall be weak. In the horizontal plane, two pairs of orthogonal
contra-propagating laser beams red detuned with respect to the
$^1\!S_0\rightarrow\,^3\!P_1$ narrow transition are turned on (see
Fig. \ref{fig1}). This set-up realizes the proposed 2D
gravitational-like interaction. In order to avoid any spatial
dependency of the quasi-resonant laser beams detuning, the dipole trap
wavelength is tuned on the so-called "magic" wavelength which is
$\lambda\sim 900\,\textrm{nm}$ for our particular case
\cite{katori1999}. Importantly, the cold cloud has a horizontal
pancake shape. This strong shape asymmetry is necessary to reduce the
repulsive interaction mediated by the multiple
scattering~\cite{Walker1990}. In this geometry scattered photons are
likely to escape the cloud through the transverse direction. Similar
requirements were successfully implemented in the one dimensional
case~\cite{Chalony2013}. They also prevent the generalization of this
method to three dimensional gravitational systems.

We have to check that the regime where the collapse should take place
is within reach of current experimental techniques.  For this order of
magnitude computation, we use a cold cloud with $N=2\times 10^6$ atoms
at a temperature of $T=2\,\mu \textrm{K}$. The power of the dipole
trap laser beams is $3$W and its waist $300\,\mu \textrm{m}$.  The
trap depth is $\frac{k_b}{2}\times 10\,\mu \textrm{K}$. The transverse
cloud size, frozen by the standing wave trapping, is set to
$L_{\bot}=5\,\mu \textrm{m}$, whereas the equilibrium longitudinal
thermal distribution in the dipole trap and without the quasi resonant
laser beams is $L=100\,\mu \textrm{m}$.  Modeling the shadow effect by
a long-range gravitation-like force requires a weak absorption of the
laser beams, \emph{i.e.} a low optical depth $b<1$; it corresponds
here, to a frequency detuning of $|\delta|>5.7\Gamma$.  In this range,
the minimal dimensionless temperature that should be reachable is
around $\Theta \sim 0.07$. This is below the theoretical threshold for
collapse, thus the expected phase transition should be observable. One
has to make sure however that the weak absorption limit is fulfilled,
and thus the model is valid, for a large range of spatial
density. Indeed we do not expect strictly speaking a collapse of the
atomic cloud since above a certain density necessarily $b>1$. In this
latter case, the shadow force becomes short range and the size of the
cloud should remain finite.  The modeling we have used also requires
a low saturation.  For this computation, we have assumed a quasi
resonant laser intensity $I=14I_s$ (where $I_s=3\,\mu W/\textrm{cm}^{2}$
is the saturation intensity).  It corresponds to a saturation
parameter $s=\frac{I/I_s}{4(\delta/\Gamma)^2+1}<0.1$.  Thus the low
saturation approximation is fulfilled.

Finally one notices that the experiment can be in principle performed
using more standard alkali setup with broad
transitions rather than the narrow intercombination
line of Strontium. However, it is expected to be technically more
challenging with the formers because "magic" wavelengths are usually
more difficult to access \cite{arora2007} and the dipole trap laser
should have a much larger power to maintain the higher temperature
gas. Moreover, temperatures close to the Fermi temperature have been
reported for laser cooling of the Strontium87 isotope in a 3D trap
\cite{mukaiyama2003recoil}. It seems reasonable to believe that the
action of laser cooling in combination with the long range attractive
force in the 2D trap might bring the gas closer or even below the
degeneracy temperature. If such a condition is fulfilled, like for a
white dwarf, the Pauli pressure should play a role in the short range
stabilization of the gas in the collapse phase. The interplay between
the non equilibrium collapse phase and the Pauli pressure remains an
open question which should be addressed in a forthcoming publication.

\emph{Conclusion-} This work paves the way for the experimental
observation of a non equilibrium collapse phase transition, driven by
a long range interaction force. All the characteristic features on the
density and the current, should be observable using the current
in-situ or time-of-flight imaging technics. This work also opens the
door to outstanding theoretical questions: how could one prove the
conjectured collapse? Beyond the entropic computation done here, what
could be the tools for such a task?  How could one develop a better
numerical scheme when the transition is approached?

\emph{Acknowledgments}: We thank Y. Brenier, M. Chalony, D. Chiron,
T. Goudon, M. Hauray, P.E. Jabin and A. Olivetti for useful
discussions. This work was partly supported by the ANR 09-JCJC-009401
INTERLOP project and the CNPq (National Council for Scientific
Development, Brazil).

\end{document}